
\input phyzzx.tex

\def\qgg {\Delta_{\hat g}}
\def\d {\partial}

\def\l {\lambda}
\def\a {\alpha}
\def\g {\hat g}

\def\fl {\phi_{\picc{L}}}
\def\fw {\phi_{\picc{W}}}
\def\Ql {Q_{\picc{L}}}
\def\Qw {Q_{\picc{W}}}
\def\b {\beta}
\def\G {\Gamma}
\def\F {\Phi}
\def\picc {\scriptscriptstyle}

\def\e {\epsilon}

\titlepage
\title{A generalized model for two dimensional quantum gravity and dynamics
of random surfaces for $d>$1}

\author {M. Martellini\foot{On leave of absence from Dipartimento di Fisica,
Universit\'a
di Milano, Milano, Italy and I.N.F.N., Sezione di Pavia, Italy}}
\address{I.N.F.N., Sezione di Roma "La Sapienza", Roma, Italy}
\author {M. Spreafico}
\address{Dipartimento di Matematica, Universit\`a di Milano, Milano
and I.N.F.N., Sezione di Milano, Italy}
\author {K. Yoshida}
\address{Dipartimento di Fisica, Universit\`a di Roma, Roma and
I. N. F. N., Sezione di Roma, Italy}
\centerline{\bf In memory of Giuseppe Occhialini}
\abstract{The possible interpretations of a new continuum model
for the
two-dimensional quantum gravity for $d>1$ ($d$=matter central charge),
obtained by carefully treating
both diffeomorphism and Weyl symmetries, are discussed. In particular we note
that an effective field theory is achieved in low energy (large area)
expansion,
that may represent smooth self-avoiding random surfaces embedded
in a $d$-dimensional flat space-time for arbitrary $d$. Moreover the values of
some critical exponents are computed, that are in agreement with some
recent numerical results.}
\endpage

\pagenumber=1
\vglue0.4cm
{\bf 1. Introduction}
\vglue0.6cm
In ref. [1]
we have argued that the fixing of the
two dimensional diffeomorphisms, by the so-called
conformal gauge
($g_{\mu\nu}\to e^{\a\fl} \g_{\mu\nu}$), as well as of the Weyl structure, by
the gauge
$R(g)=R_0$ ($R(g)$ is the curvature scalar relative to the metric $g$),
produces the following two scalar Liouville like action,
up to corrections which are negligible for large intrinsic area $A$,
$$
\eqalign{
&S[\fl,\fw;\g] = S_0 + {1\over 2\l} \int d^2\xi \sqrt{\g} (R(e^{\a\fl}\g) -
R_0)^2 + O({1\over A}) \cr
&S_0[\fl,\fw;\g] = {1\over 8\pi} \int d^2\xi \sqrt{\g}
[-M_{ij}\Phi^i\qgg\Phi^j-Q_i\Phi^i R_{\g}]\cr}
\eqn\action
$$
where
$$
\eqalign{
&\Phi^i\equiv (\fl,\fw)\cr
&Q_i\equiv(\Ql,\Qw)\cr
&M_{ij}\equiv\left(\matrix{1&B\cr
B&-1\cr}\right)\cr}
$$
and we use here the same notations as in ref. [1].

This model, which is a sort of generalization of the DDK one [2],
was constructed in [1] by requiring conformal invariance and the "boundary
condition" that for $d\to 1^+$ the action
\action\ should reproduce the DDK model.
Ignoring the last terms in eq. \action, at least to calculate
the genus $h$ string susceptibility $\G_{h}$, we
have shown that for $d>1$ this model gives a totally new physics.
Indeed in order to calculate $\G_{h}$ defined by the
scaling relation
$$
Z({A})\sim K A^{\G_{h}-3}e^{-\mu A}, ~~~A\to\infty
\eqn\scaling
$$
it is sufficient to consider only the large $A$ behaviour.
This procedure is consistent with the recent result of Kawai and Nakayama
(KN) [3]
which show that, for large $A$, the $R^2$-term does not
influence the leading behaviour.
Furthermore notice that the non-analytic scaling breaking term found by KN,
namely $exp\left(-{32\pi^2 (1-h)\over A}\right)$, is exactly cancelled
in \scaling\ by the contribution coming from the fixed curvature
scalar $R_0$ in \action\ by using
the Gauss-Bonet identity.
Therefore we must think of \action\ as a sort of "low-energy effective field
theory"
depending on two free parameters $B$ and $d$.

{}From now on, our basic strategy is to regard $S_0$ in \action\
as the starting point to
construct the fixed area partition function
$$
\eqalign{
Z(A) =& \int D_{\hat g}X D_{\hat g}\fl D_{\hat g}\fw
  D_{\hat g}b D_{\hat g}c  d\tau\cr
& e^{-S_{\picc M}[X;\hat g]-S_{\picc GH}[b,c;\hat g]-S_0 [\fl,\fw;\hat g]}
\delta\left(\int d^2 x \sqrt{g} -
A\right) e^{-\mu A}\cr}
\eqn\partfunz
$$
where $S_{\picc M}$ is the usual Polyakov action associated with the
immersion maps $X^a:\Sigma_h\to R^d$,$a=1,~...,~d$, implying the
matter central charge $c_{\picc M}=d$.
The main result of [1] was that the genus $h$ string susceptibility
associated with \partfunz\ is
$$
\Gamma_h={2 (1-h)\over \sqrt{1+B^2}}
{B\sqrt{1+Bd}-\sqrt{(1+B^2)(25+(B-1)d)}\over
\sqrt{25+(B-1)d}-\sqrt{1+(B-1)d}} +2
\eqn\stringsusc
$$

$\G_{h}$ is real for $d>1$ if we assume that $B\geq B_c=O(1)$.
Notice that for $B=0$, \stringsusc\ is the DDK string susceptibility.
This fact implies that we may regard \partfunz\ as an off-critical string
theory for $d\geq 1$, which extends the DDK model (that exists only for
$d\leq 1$) and belongs to the same universality class as DDK at $d=1$.

The first (almost trivial) consequence of \scaling\ and \partfunz\
is that, even for $c_{\picc M}=d>1$, the mean (intrinsic) area is
a linear function of the genus $h$. Indeed by defining
$$
Z_{h}(\mu)=\int dA Z_h(A)\sim\mu^{-(\G_h -2)}
$$
we have that
$$
\eqalign{
&<A>_h = - {\d\over\d\mu} ln Z_h (\mu)\sim {\G_h -2\over \mu}\cr
&{\d\G_h\over\d h} =
-{2\over \sqrt{1+B^2}}
{B\sqrt{1+Bd}-\sqrt{(1+B^2)(25+(B-1)d)}\over
\sqrt{25+(B-1)d}-\sqrt{1+(B-1)d}}\cr}
\eqn\meanarea
$$

The linear behaviour \meanarea\ is consistent with some recent numerical
simulations
[4] [5] [6].
We shall see also that the values of $\G_h$ and its slope ${\d\G_h\over\d h}$
calculated according to eqs. \stringsusc\ and \meanarea\ agree well with
the numerical results obtained in [4] [5] [6].

In this letter we obtain two main results:

- If we consider \partfunz\ as a generalized DDK model, we get a
closer information on the value of $B$ by studying the dynamics, i.e.
the associated tachyon scattering amplitudes.
It turns out that $B\sim B_c\equiv 1-{1\over d}$. Notice that this choice
automatically guarantees our boundary condition for $d\to 1^+$. This analysis
is developed in section 2.

- As it is well known, in the so-called dynamical triangulated
approach of the two-dimensional quantum gravity, there are no obstruction
in the construction of a random surface model for $d>1$
[7].
Therefore, it is of interest to ask if one can relate for $d>1$ the continuum
theory described by the model \partfunz\ with some underlying random surface
model.
We shall show in section 3, that if we regard $B$ as
an effective coupling parameter independent from $d$, our model may describe
a phase of smooth, self-avoiding random surface embedded in $R^d$
for $B\sim B_H\equiv 1$, with a positive $\G$.

In the following we shall assume the planar topology, i.e. $h=0$.
\vglue0.4cm
{\bf 2. Tachyon scattering amplitudes}
\vglue0.6cm
We proceed to study the tachyon scattering amplitudes of our model coupled
to a conformal matter represented by the following action
$$
S_{\picc M}= {1\over 8\pi} \int d^2 x \sqrt{\g} [- X^a \qgg X_a ]
$$

Consistently with our interpretation of the area operator, the tachyon vertex
operators
$V_k=e^{ik_a X^a}$ get a gravitational dressing by the Liouville field
$\F^1=\fl$ alone
$$
T_k = e^{i k_a X^a + \b_1 \F^1}~~~~~~~~(k_a=k_{a'}~\forall a,a')
$$
where the parameter $\b_1=\b(k)$ is defined by conformal invariance
($\Delta[T_k]=1$)
$$
\b_{\pm}(B,k)=-{\sqrt{1+B^2}\over 2\sqrt{3}} \left [ \sqrt{25+d (B-1)}\mp
\sqrt{1+d (B-1) +12 d k^2}\right]
$$

As a result we can express the tachyon scattering amplitudes over a
fixed background geometry with planar topology
as follows
$$
\eqalign{
&{\cal A}_n(k_1, ..., k_n):=<T_{k_1}\dot ... \dot T_{k_n}>=\cr
&={2\sqrt{\pi}\over \a_1} A^{-1-s} \int D_{\g} \F' D_{\g} X'
e^{-I[\F';\g] -S_{\picc M}[X';\g]}
\left [\int d x^2 \sqrt{\g}
e^{\a_1\F'^1(x)}
\right]^s T_{k_1} ... T_{k_n}\cr}
\eqn\corr
$$
where $\F'$ and $X'$ represent the non
zero modes of the fields, $I$ is the quadratic part of the action $S_0$ defined
in eq. \action. In \corr\
$s$ is given by
$$
s= -{Q_1\over \a_1} - {1\over \a_1}\sum_{i=1}^n \b(k_i)
\eqn\esse
$$
$$
\eqalign{
&Q_1=\Ql\equiv -{1\over \sqrt{3}}\left[B\sqrt{1+Bd}
-\sqrt{(1+B^2)(25+(B-1)d)}\right]\cr
&\a_1=-{\sqrt{1+B^2}\over 2\sqrt{3}} \left[\sqrt{25+(B-1)d}-\sqrt{1+(B-1)d}
\right]\cr}
$$

The charge neutrality of the matter sector requires
$$
\sum_{i=1}^n k_i =0
\eqn\conscarica
$$

In case of integer $s$ we can reduce \corr\ to the following multiple integral
$$
\eqalign{
&{\cal A}_n={2\sqrt\pi\over \a_1} A^{-1-s} \cr
&\prod_{i=1}^n \prod_{j=1}^s
\int d^2 x_i d^2 t_j |x_i-t_j|^{-{2\over 1+B^2} \a_1 \b_i}
\prod_{i'<i} |x_{i'}-x_i|^{-{2\over 1+B^2}\b_{i'} \b_i +2 k_{i'} k_i}
\prod_{j'<j} |t_{j'}-t_j|^{-{2\over 1+B^2}\a_1^2}\cr}
$$

We must now find what are the values of $B$ for which
these correlators have a "good behaviour".
In ref. [8]
we calculated the three-points correlation function and, studying
its convergence conditions, we got an indication that
$B\geq B_c\equiv 1-{1\over d}$. Furthermore $B=B_c$ is the minimum value
that ensures the reality of $\G$. We also found that the allowed integer
values for $s$ in \corr\ and \esse, were $1$ and $s=2$. Applying similar
technique to the four-points function
$$
\eqalign{
&{\cal A}_4(k_1, k_2, k_3, k_4) =\cr
&= {2\sqrt\pi\over \a_1} A^{-1-s}
\int d^2 x |x|^{2a'} |1-x|^{2b'}
\prod_{i=1}^s \int d^2 t_i |t_i|^{2a} |1-t_i|^{2b} |x-t|^{2p}
\prod_{i'<i} |t_{i'}-t_i|^{4c}\cr}
$$
where
$$
\eqalign{
&a=-{1\over 1+B^2} \a_1\b(k_1);~~~~~~~b=-{1\over 1+B^2} \a_1\b(k_3)\cr
&p=-{1\over 1+B^2} \a_1\b(k_4)\cr
&a'=k_1 k_4-{1\over 1+B^2}\b(k_1)\b(k_4);~~~~b'=k_3 k_4-{1\over 1+B^2}
\b(k_3) \b(k_4)\cr
&c=-{1\over 2(1+B^2)} \a_1^2\cr}
$$
we find that it never converges for $s=2$. Instead it converges
for $s=1$ under the following conditions
$$
\eqalign{
&Re ~a,b,a',b'>-1\cr
&-Min\left[ {1\over 2}; Re {a+a'\over 2}+1; Re {b+b'\over 2}+1\right]
<Re{p\over 2}<\cr
&<-Max\left[ {1\over 2} Re(a+b+1); {1\over 2} Re (a'+b'+1);
{1\over 2} Re (a+b+a'+b'+1) +1\right]\cr}
\eqn\pl
$$

Remembering \esse\ and \conscarica\ and the above conditions,
we find "experimentally" that for fixed $d$ and for
$B\in [1-{1\over d},1-{1\over d}+\e],\e\sim 0$
there exists a region in the $\{k_1,k_2, k_3, k_4\}$
$4$-dimensional momentum-space where $A_4$ converges.
Moreover
we note that a strong simplification arises in the
form of the parameters when $B=B_c\equiv 1-{1\over d}$,
since in this case they becomes linear
functions of (the moduli of) the momenta. In fact substituting $B=B_c$ in the
definitions we find that
$$
\eqalign{
&Q_1(d)=2\sqrt{2} {\sqrt{2d^2+1-2d}\over d} -\left( 1-{1\over d}\right)
\sqrt{d\over 3}\cr
&\a_1(d)=-\sqrt{2} {\sqrt{2d^2+1-2d}\over d}\cr
&\beta(d,k)={\sqrt{2d^2+1-2d}\over d}(\sqrt{d}|k|-\sqrt{2})\cr
&a,b,p=\sqrt{2d}|k_{1,3,4}|-2\cr
&a',b'=\sqrt{2d} (|k_{1,3}|+|k_4|)-3|k_{1,3} k_4|-2\cr}
$$

We end this section by outlining the argument that allow us to identify
the region ("window") of convergence of ${\cal A}_4$ (for $s=1$).
We note that, because of \esse\ and \conscarica, there are only two independent
momenta, say $k_1$ and $k_4$. Indeed by choosing the kinematic region
$k_1, k_3, k_4 >0$ and $k_2 <0$, we get
$$
\eqalign{
&k_2= -{1\over\sqrt{2d}} \left( {(d-1) \sqrt{d}\over\sqrt{12 d^2-12d+6}}+3
\right)\cr
&k_3= -k_1+|k_2|-k_4\cr}
$$

Using the inequality \pl\ (with $B=B_c$), we find the condition of convergence
in the ($k_1$, $k_4$) plane, that is
$$
\eqalign{
&k_1> {1\over\sqrt{2d}}\cr
&{1\over\sqrt{2d}}<k_4<{2\sqrt{2d}\over 4d-3}\cr}
$$

An explicit calculation of the four-points correlation function can be found in
[9].
\vglue0.4cm
{\bf 3. Planar random surface models}
\vglue0.6cm
In this section we analyse which kind of
planar random surface model may be related to the "generalized" two-dimensional
quantum gravity theory discussed in the previous section. We have already
remarked that \partfunz, regarded as a low-energy (large-area) effective field
theory, in the planar limit, depends on the "coupling constant" $B$ and
on the dimension $d$ of the target space. Therefore
if we understand the model \partfunz\
as a possible "phase" of the two-dimensional quantum gravity for $d>1$
(complementary to that described by the DDK model for $d<1$), $B$ is
an effective parameter and the geometrical complexity of the associated
planar random surfaces changes by varying $B$ in the ranges determinated
by some "critical" values, as we shall show below.
For this purpose, we consider, together with the string susceptibility
$\G=\G(B,d)$ defined in eq. \stringsusc, the value of the Hausdorff dimension
$d_H$ as a function of $B$ and $d$.

We compute $d_H(B,d)$ following the method of ref. [10].
The key formula is given by the relation
between the mean square size of the embedded surface and the
two-points correlation function in momenta space, which in our model is a
particular case of eq. \corr. The result for large $A$ reads
$$
d_H(B,d)=2({\b_+(B,d,k^2)+\b_-(B,d,k^2)\over \a(B,d)}-2)|_{k^2=0}=
\sqrt{25 +d(B-1)\over 1+d(B-1)}-1
\eqn\dimh
$$

In the following we distinguish two classes of planar random surface models
associated to \partfunz.

{\it (a) Off-critical string model for $d\geq 1$}

In section 2 the model \partfunz\ is seen as an off-critical
string with $d\geq 1$, which must be in the same "universality class"
of the DDK model at $d=1$. This implies in particular that at $d=1$
it must have the same
string susceptibility and Hausdorff dimension as the DDK model.
Furthermore, analyticity
conditions on the $n$-points correlation functions require $B=B_c+\e,\e\sim 0$.
Thus we get, for $d\geq 1$ ($\e\sim 0$)
$$
\eqalign{
&\G(d)=\G_c(d)+o(\e)\cr
&\G_c(d)={1\over \sqrt{6}} \sqrt{{d\over 2d^2-2d+1}}(d-1)\cr
&{\d\G_h(B_c,d)\over\d h}= 2-\G_c(d)\cr}
\eqn\undici
$$
and
$$
d_H(d)\sim \sqrt{{24\over d}} \e^{-{1\over 2}}
$$

The plot of $\G_c(d)$ is given
in \fig{Behaviour of $\G_c(d)$ for arbitrary values of $d$.}.
Note that the values of $\G_c(d)$ for $1<d\leq 4$ fit rather well with the
recent numerical data obtained in ref.s [4] [5] [6]. In particular,
we get $\G_c(d=3)= 0.392\sim 0.4$ which must be compared with the planar
exponents $b=2-\G_c(3)$ and $\gamma=\G_c(3)$ of [4] and [6] respectively.
Indeed, we find $b\sim 1.6$, which is consistent (within the large numerical
errors) with the upper bound of $b$ ($b\sim 0.93$) given by Caselle et al.
[4] in the case of the "smaller" lattice.
The correspondence appears to improve if one uses the new efficient algorithm
for sampling of random surfaces in the Monte Carlo simulations used by Ambjorn
et al. [6].
These last named authors analyse the fractal structure of the two-dimensional
quantum gravity coupled to conformal matter with central charge $c_{\picc M}$,
by measuring the distribution of the so-called baby universes and the numerical
algorithm based on the cutting and pasting of the baby universes.
In [6], one uses several fits to compute the critical exponent $\gamma$.
However, in order to compare $\gamma(c_{\picc M})$ with our $\G_c(c_{\picc M}
=d)$, one must choose in [6] the fit which gives $\gamma(c_{\picc M})\sim 0$
as $c_{\picc M}\to 0$.
This is the "fit({\it b})" in ref. [6] and, for which this one, one finds
$\gamma_{\it b}(3)=0.53$ (11) [6], a value very close to our result
$\G_c(3)\sim 0.4$.
As a final check of eq. \undici, note also that ${\d\G_h (B_c,d=3)\over\d h}
=1.608\sim 1.6$ and this value agrees well with the slope found in [5].
The authors of ref. [5] find, by using a body centered cubic lattice and the
Swendsen-Wang (SW) Monte Carlo algorithm, a slope $1.25\pm 0.1$.

As for the geometrical complexity of the random surface associated
to the behaviour summarized by eq. \undici, we have mentioned in the
introduction the equivalence between
our model and the higher curvature ($R^2$)
KN model obtained by adding the
$R^2$ term to the Gibbs measure of \partfunz.
This observation
suggests that the surfaces involved avoid the branched-polymer
configurations by the KN mechanism. However, the fact that the
Hausdorff dimension is very large with respect to the physical dimension
$d$, indicates that these surfaces are not self-avoiding.

{\it (b) Self-avoiding planar random surfaces}

Alternatively we may think of \partfunz\ as an effective field description
of the phase of the two-dimensional quantum gravity
for $d>1$. Thus $B$ is an effective
parameter and we may ask whether there exists a value
$B_H$ of $B$ such that
$$
\eqalign{
&d_H(B_H,d)=k\cr
&\G(B_H,d)>0\cr}
\eqn\equazione
$$
for some open interval of $d>1$, like for example $d\in (1,24]$, with
$k$ a fixed positive finite constant. Then we have self-avoiding (SA)
planar surfaces if we choose $k<d$.
{}From eq. \equazione, we get that
$$
B_H=B_H(k,d)=1+{(6+k)(4-k)\over kd(k+2)}
\eqn\ultima
$$

We have studied the two most relevant cases:

{\it i}) k=1. Then by construction $d>d_H$ for any $d\in (1,24]$ (hence SA
regime)
and by \ultima\ $B_H=1+{7\over d}$; the plot of $\G(1+{7\over d},d)$ is
given in
\fig{Behaviour of $\G(1+{7\over d},d)$ for arbitrary values of $d$.}.
Notice that $\G(d=1)\cong 0.105$. Therefore this SA random surface
model is not in the same
universality class of DDK, but for say $d=3$ we have $\G(d=3)\cong 0.246$.
So we conjecture that our model may supply a microscopic field theory
description for the interface magnetization domains of the 3D-Ising
model [4].

{\it ii}) k=4. Now by \ultima\
$B_H=1$ and the plot of $\G(1,d)$ for $d\in(1,24]$ is given in
\fig{Behaviour of $\G(1,d)$ for arbitrary values of $d$.}. In particular
$\G(1,d=1)=0$.
Therefore, in this limiting case, we get the same asymptotic partition function
for large $A$ as the DDK model, but with a finite Hausdorff dimension.
This class of planar random surfaces is SA only for $d>4$.

Let us conclude with an observation. We have shown above that our
model may describe the statistical mechanics of planar random surfaces, in
a SA-phase or not, if $B$ is treated as an effective coupling parameter.
This is consistent with our approximation of neglecting the potential
interaction term in \action\ and of regarding \partfunz\ as a
sort of low-energy effective field theory.
Of course we do not know up to now
the complete quantum field theory leading to
\partfunz, and hence we can not say whether or not the consistency
conditions applied
to the full theory are brought to fix the value of the parameter $B$.
As we have shown in [1] this does not happen at the level of the
effective field theory model \partfunz.
\vglue0.4cm
{\bf Acknowledgements}
\vglue0.6cm
We thank M. Caselle, M. Ghiozzi and E. Marinari for a series of illuminating
discussions. The work is partially supported by Ministero della
Ricerca Scientifica e dell'Universit\`a 40\%.
\vglue0.4cm
{\bf References}
\vglue0.6cm
\parindent 0.0pt
{[1] M. Martellini, M. Spreafico and K. Yoshida, Mod. Phys. Lett. A7 (1992)
1667;}

{[2] J. Distler and H. Kawai, Nucl. Phys. B321 (1989) 509;}

{[3] H. Kawai and R. Nakayama, Phys. Lett. B306 (1993) 224;}

{[4] M. Caselle, F. Gliozzi and S. Vinti,
"Condensation of handles in the
interface of the 3D Ising model", DFTT preprint, DFTT-12/93;}

{[5] V. S. Dotsenko,
M. Picco, P. Windey, G. Harris, E. Marinari and E. Martinec, "The
phenomenology of strings and clusters in the 3D Ising model", Univ. Roma
preprint (1993);}

{[6] J. Ambjorn and G. Thorleifsson, "A universal
fractal structure of 2D quantum gravity for $c>1$", NBI-HE-93-64 preprint;}

{[7] J. Ambjorn, B. Durhuus, J. Frohlich and P. Orland, Nuc. Phys.
B270 [FS16] (1986) 457;}

{[8] M. Martellini, M. Spreafico and K. Yoshida, in "String
Theory, Quantum Gravity and the Unification of the Fundamental Interactions",
ed.s M. Bianchi, F. Fucito, E. Marinari and A. Sagnotti, World Scientific
(Singapore 1993);}

{[9] M. Martellini, M. Spreafico and K. Yoshida, "A Continuum Approach to
2D-Quantum Gravity for $c>1$", Milano/Roma preprint (1993);}

{[10] J. Distler, Z. Hlousek and H. Kawai,
Int. J. Mod. Phys. A5 (1990) 1093;}
\vfill
\endpage
\figout
\vfill
\endpage
\end